# Strong Interfacial Exchange Field in 2D Material/Magnetic-Insulator Heterostructures: Graphene/EuS

**Authors:** Peng Wei[1,2,*], Sunwoo Lee[3,4], Florian Lemaitre[3,5,a], Lucas Pinel[3,5,b], Davide Cutaia[3,6,c], Wujoon Cha[7], Ferhat Katmis[1,2], Yu Zhu[3], Donald Heiman,[8] James Hone,[7] Jagadeesh S. Moodera[1,2], Ching-Tzu Chen[3,†]

**Affiliations:**

1. Francis Bitter Magnet Laboratory, Massachusetts Institute of Technology, Cambridge, MA 02139, United States

2. Department of Physics, Massachusetts Institute of Technology, Cambridge, MA 02139, United States

3. IBM TJ Watson Research Center, Yorktown Heights, NY 10598, United States

4. Electrical Engineering Department, Columbia University, New York, NY 10027, United States

5. Institut polytechnique de Grenoble, F38031 Grenoble Cedex 1 - France

6. Politecnico di Torino, Turin 10129, Italy

7. Mechanical Engineering Department, Columbia University, New York, NY 10027, United States

8. Department of Physics, Northeastern University, Boston, MA 02115, United States

a. Present address: Department of Electrical Engineering, Eindhoven University of Technology, 5612 AZ, Eindhoven, The Netherlands

b. Present address: Department of Electronic & Electrical Engineering, The University of Sheffield, Mappin Street, Sheffield S1 3JD, United Kingdom

c. Present address: IBM Zurich Research Laboratory, Säumerstrasse 4, CH- 8803, Rüschlikon, Switzerland

* pwei@mit.edu

† cchen3@us.ibm.com

Exploiting 2D materials for spintronic applications can potentially realize next-generation devices featuring low-power consumption and quantum operation capability.[1-3] The magnetic exchange field (MEF) induced by an adjacent magnetic insulator enables efficient control of local spin generation and spin modulation in 2D devices without compromising the delicate material structures.[4,5] Using graphene as a prototypical 2D system, we demonstrate that its coupling to the model magnetic insulator (EuS) produces a substantial MEF (> 14 T) with potential to reach hundreds of Tesla, which leads to orders-of-magnitude enhancement in the spin signal originated from Zeeman spin-Hall effect. Furthermore, the new ferromagnetic ground state of Dirac electrons resulting from the strong MEF may give rise to quantized spin-polarized edge transport. The MEF effect shown in our graphene/EuS devices therefore provides a key functionality for future spin logic and memory devices based on emerging 2D materials in classical and quantum information processing.

**One Sentence Summary:** Experimental demonstration of intense (> 14 T) magnetic exchange field (MEF), the induced spin-polarization, and the unconventional ferromagnetic ground state in a prototypical 2D material magnetic heterostructure: graphene/EuS

Magnetic exchange field in magnetic multilayers can potentially reach tens or even hundreds of Tesla.[6] The single-atomic-layer (2D) materials, such as graphene, mono-layer $WS_2$ etc., is expected to experience the strongest MEF in heterostructures with magnetic insulators due to the short-range nature of magnetic exchange coupling.[4] 2D material/magnetic insulator heterostructures enable local spin modulation by magnetic gates,[4,5,7] and the realization of efficient spin generation for spintronic applications.[8,9]

As a proof of concept, here we demonstrate substantial MEF and spin polarization in CVD graphene/EuS heterostructures. We have chosen EuS as a model magnetic insulator because of its wide band-gap (1.65 eV), large exchange coupling $J \sim 10$ meV, and large magnetic moment per Eu ion $\langle S_z \rangle \sim 7~\mu_B$,[10] yielding large estimated exchange splitting $\Delta \propto J\langle S_z \rangle$ in graphene.[4,5] EuS has also been shown to spin-polarize quasiparticles in materials including superconductors and topological insulators.[6,11] The strength of the MEF depends critically on the interface and EuS quality,[12,13] which we optimize with an in-situ cleaning and synthesis process (Methods and Fig. 1a). In contrast to other means, such as defect- or adatom-induced spin polarization,[14,15] depositing insulating EuS well preserves graphene's chemical bonding, confirmed by Raman spectroscopy (Fig. 1b). The slight enhancement of the Raman *D*-peak suggests that EuS deposition introduces a small number of scattering centers, as reflected in the decrease in electronic mobility. [See S1 in Supplementary Information (SI).] However, most graphene/EuS devices develop Schubnikov de Haas (SdH) oscillations in field $\geq 2$ T at 4.2 K, and the lowest quantum Hall plateaus locate exactly at $2e^2/h$ (Fig. S5-1), indicative of high graphene quality and well-preserved Dirac band structure.

We utilize Zeeman spin-Hall effect (ZSHE) to probe the MEF in graphene which splits the Dirac cone via Zeeman effect and generates electron- and hole-like carriers with opposite spins near the Dirac point $V_D$ (Fig. 2a right panel).[8,9] Under a Lorentz force, these electrons and holes propagate in opposite directions, giving rise to a pure spin current and non-local voltage $V_{nl}$ (Fig. 2a left panel). We measure the non-local resistance $R_{nl}$ of ZSHE using the device configuration in Fig. 2a (see Methods). Figure 2b plots $R_{nl}$ as a function of the back-gate voltage $V_g$ under a series of applied field $\mu_0 H$ for a graphene/EuS device. The $R_{nl}$ peak at $V_g = V_D$ can be described by:[8,9,16,17]

$$R_{nl,D} \propto \frac{1}{\rho_{xx}} \left( E_Z \frac{\partial \rho_{xy}}{\partial \mu} \right)^2 \bigg|_{\mu_D} \qquad (1)$$

where $\rho_{xx}$ ($\rho_{xy}$) is the longitudinal (Hall) resistivity, $E_Z$ is the Zeeman splitting energy, and $\mu$ is the chemical potential. As shown in Fig. 2c, $R_{nl,D}$ in graphene/EuS easily dwarfs that in pristine graphene by a factor of ~ 8, revealing a substantial $E_Z$ enhancement. Moreover, after prolonged air exposure (~ 1 month), $R_{nl,D}$ decreases significantly due to degraded EuS magnetic properties by oxidation.

We further confirm the effect of EuS via correlated studies of transport and magnetization. Fig. 2d shows that both $R_{nl,D}$ and magnetization $M$ in graphene/EuS rise abruptly as $T$ drops below $T_C$ (~ 16 K) of EuS. On the other hand, the weak $T$ dependence of $R_{nl,D}$ in graphene/AlO$_x$ (Fig. 2d inset) is consistent with that in graphene/EuS above $T_C$. The $R_{nl,D}$ at $T > T_C$ reflects the contributions from the applied-field induced ZSHE,[7,9] and the paramagnetic response of EuS at high field (e.g. 3.5 T).[10,11] Nevertheless, the MEF induced $R_{nl,D}$ dominates in graphene/EuS when $T < T_C$. Lastly, Fig. 2e demonstrates that in comparison with the longitudinal resistance $R_{xx}$, $R_{nl}$ shows a 10-fold larger change in magnitude and a much narrower peak under finite field, thereby excluding the Ohmic contribution. In contrast, the small zero-field $R_{nl}$ *does* scale with $R_{xx}$, providing a quantitative estimate of the Ohmic contribution (See S2.1 in SI). Other extrinsic effects are further discussed and ruled out (see S2).

To estimate the $E_Z$ enhanced by the MEF, we compare the field dependence of $R_{nl,D}$ in graphene/EuS with that in graphene/AlO$_x$ (Fig. 3a) fabricated from the same batch of CVD graphene. The normalized $R_{nl,D}(\mu_0 H)/R_{nl,D}(3.8\ T)$ variation in graphene/EuS is almost two orders of magnitude larger, indicative of strong contribution from EuS. $R_{nl,D}(3.8\ T)$ is chosen as the reference since ZSHE dominates the non-local signal at large field.

According to Eq. (1) and prior reports,[9,16,17] $R_{nl,D}$ can be recast as

$$R_{nl,D} = R_0 + \beta(\mu_0 H) \cdot E_Z^{\ 2} \qquad (2).$$

Here $R_0$ accounts for the zero-field non-local signal from extrinsic sources (see S2 in SI), the parameter $\beta$ represents the orbital-field effect manifested by $\rho_{xx}$ and $\rho_{xy}$, and $E_Z = g\mu_B B_Z = g\mu_B(B_{exc} + \mu_0 H)$ where $B_{exc}$ is the MEF. We further define the parameter $\alpha$:

$$\alpha^2(\mu_0 H) \equiv \frac{\beta(\mu_0 H)}{\beta(\mu_0 H_0)} = \left(\frac{E_{Z0}}{E_Z}\right)^2 \cdot \left[\frac{R_{nl,D}(\mu_0 H) - R_0}{R_{nl,D}(\mu_0 H_0) - R_0}\right] \qquad (3),$$

where $E_{z0}$ denotes the Zeeman energy at the reference field $\mu_0 H_0$. Given $H_0$, deriving $\alpha(\mu_0 H)$ of graphene/AlO$_x$ is straightforward because $E_z$ is solely determined by $\mu_0 H$. The inset of Fig. 3(b) shows the calculated $\alpha$ using $\mu_0 H_0 = 1$ T, a proper reference field as we will explain below.

To derive $\alpha$ of graphene/EuS, we note that according to the theory of ZSHE,[9,17] $\alpha$ depends on sample mobility, while other sample-dependents terms (including spin relaxation length, density of thermally activated carriers and Fermi velocity) cancel out (see S3 in SI). The mobility difference between our graphene/EuS and graphene/AlO$_x$ samples is ~25% (see S1 in SI), which would only yield a ~10% correction to α (see S3 in SI). Since ~10% difference is small, for an order-of-magnitude estimate of the MEF, we adopt the $\alpha$ value of graphene/AlO$_x$ for graphene/EuS as an approximation. We then evaluate $E_Z$ in graphene/EuS using $E_Z = \frac{E_{Z0}}{\alpha}\sqrt{\frac{R_{nl,D}(\mu_0 H) - R_0}{R_{nl,D}(\mu_0 H_0) - R_0}}$.

To obtain the lower bound of $E_Z$, we approximate $E_{Z0} \approx g\mu_B\mu_0 H_0$, ignoring the $B_{exc}$ contribution. This constrains us to use a small $H_0$ such that $B_{exc}$ is small. Meanwhile, $H_0$ should be high enough to ensure that $R_{nl,D}(\mu_0 H_0) - R_0$ is much larger than noise. Therefore we choose $H_0 \approx 1$ T, above and close to the onset of $R_{nl,D}$ (Fig. 3a). The calculated *lower-bound* estimate of $E_Z$ is plotted against the applied field in Fig. 3b. We further take the free electron gyromagnetic factor $g = 2$ to estimate $B_Z$ and find that it reaches a significant value of $> 18$ T when $\mu_0 H \sim 3.8$ T (Fig. 3b). Moreover, the spin generation efficiency of ZSHE characterized by the spin Hall angle $\theta_{SH}$ is enhanced from $\theta_{SH} \sim (0.51 \pm 0.10)$ at $\mu_0 H = 2\,T$ in pristine graphene to $\theta_{SH} \geq 1.22$ in graphene/EuS – more than a factor of 2 gain. In contrast to conventional spin injection in which spin-polarized electrons tunnel into 2D materials through a barrier,[18] the MEF directly polarizes the 2D electrons, thereby circumventing issues of pinholes and barrier breakdown.

The intense MEF also lifts the ground state degeneracy of graphene in the quantum Hall regime. In graphene/EuS, the quantum Hall regime is reached at $\mu_0 H \simeq 3.8\,T$ (Fig. 4a). The filling factor $\nu$ of the Landau level (LL) obeys: $\nu = \pm 4(|n|+1/2)$ for 2D Dirac fermions where $n$ = 0, 1, 2… is the LL index. $\nu$ can be derived from the gate voltage at each $R_{xx}$ minimum (Fig. 4a inset). By tracing the $\nu = \pm 2$ fillings with $\mu_0 H$, we find that the corresponding electron density depends only on $\mu_0 H$ but not on $B_Z$ (Fig. 4b, Fig. S2-5), implying that EuS causes negligible orbital field. Nevertheless, a close look at the $R_{xx}$ maxima reveals extra dip features that develop with increasing $\mu_0 H$, especially at the $n = 0$ LL (Fig. 4c). This peak-splitting feature indicates the lifting of LL degeneracy under large $B_Z$, which can be well fitted by simple Gaussians (Fig. 4d). The observed splitting feature is reproducible in both $R_{xx}$ and $R_{nl}$ (inset of Fig. 4d) and the associated plateau feature is also observable in the $R_{xy}$ measurements (see S5 in SI).

Another observation associated with the intense $B_Z$ is the reduction of $R_{xx,D}$, longitudinal resistance at the Dirac point. As shown in Fig. 4e, $R_{xx,D}$ initially increases with $\mu_0 H$ until it peaks at ~ 2 T and then drops continuously with field. At 3.8 T, $R_{xx,D}$ is even smaller than its zero-field value, in sharp contrast to the monotonic increase of $R_{xx,D}$ in graphene/AlO$_x$ (Fig. 4e inset) and in prior reports.[19-21] The divergent $R_{xx,D}$ is usually attributed to the gapped $v = 0$ state under a perpendicular field.[22,23] The $v = 0$ state originates from the splitting of $n = 0$ LL and can either be a valley-polarized spin singlet (Fig. 4f) or a spin-polarized valley singlet (Fig. 4g) depending on the relative strength of orbital vs. Zeeman field.[22-24] Since the valley-polarized state inevitably leads to divergent $R_{xx,D}$, the observed decrease of $R_{xx,D}$ in graphene/EuS strongly suggests the spin-polarized valley-singlet state. That $R_{xx,D}$ (3.8 T) < $R_{xx,D}$ (0 T) further indicates additional states forming at the Dirac point with increasing $B_Z$, consistent with the presence of counter-propagating edge channels in the spin-polarized $v = 0$ state (Fig. 4g) as in the quantum spin Hall effect (Fig. 4g inset).[22,25,26] Such a novel ground state had only been created under an enormous in-plane field ($\mu_0 H > 20$ T) in selected high-quality graphene devices.[22,25] Here we show that it is accessible in a much lower field using scalable CVD graphene by leveraging the MEF. Thus our approach enables other emerging phenomena, for example the quantum anomalous Hall effect in 2D materials by engineering magnetic insulators with inherently both MEF and spin-orbit-coupling,[27,28] or magnetically gated information storage/processing devices with strong spin-orbit-coupling in 2D transition metal dichalcogenides using MEF.[3,29] For industrial applications in low-power information processing, development of high quality magnetic insulators that are compatible with 2D materials and magnetic above room temperature will be highly desirable.

**Methods:**

Wafer-size (> 1cm × 1cm) monolayer graphene was grown on copper foils by standard chemical vapor deposition (CVD).[30] We transferred the CVD graphene samples to low-resistivity silicon substrates (< 0.005 Ω·cm) capped with ~ 100 nm custom-grown thermal oxide (gate dielectric for chemical potential tuning) and pre-patterned with gold alignment marks. We have employed two transfer techniques – the PMMA/PDMS stamping method and the recyclable pressure sensitive adhesive films method.[30] The latter is found to yield higher quality devices. The transferred graphene was first shaped into Hall bar geometry using electron-beam lithography (EBL) and oxygen plasma etching. The channel width of the devices ranges from 0.5 – 1 μm, and the channel length ($l$) to width ($w$) ratio is fixed to $l/w = 2, 2.5$, and 3. Then the electrodes of Ti/Pd/Au were deposited by e-beam evaporation. Lastly, to remove polymeric residues from the aforementioned fabrication processes, we have annealed the samples in high vacuum (< 1 x $10^{-6}$ Torr) at 170 – 200 ºC for ~3 hours to remove the organic residues from the e-beam resist after the device fabrication. The typical field-effect mobility of the resulting devices is about 6000 cm$^2$/V·s. Prior to forming the EuS/graphene heterostructure, the Hall bar devices were annealed in-situ at low $10^{-8}$ Torr vacuum to remove possible water molecules on graphene and any remaining organic residues immediately before the EuS deposition. Then EuS (3 ~ 7 nm) was deposited at room-temperature under $10^{-8}$ torr vacuum and capped with AlO$_x$ (15 nm) to prevent oxidation. For simplicity, in the main text, we omit this capping layer when labeling the devices. Both EuS and AlO$_x$ are grown from single compound sources. The growth of EuS is shown to preserve the graphene layer underneath by Raman spectroscopy (Fig. 1b), and the growth of AlO$_x$ is proven to preserve the EuS layer below.[11] The detailed characterizations of the heterostructure material are described in section S1 of the SI. EuS grown on graphene is confirmed to have single phase (00L) type orientation by the clear (002) peak in X-ray diffraction (XRD) data (Fig. 1c). Unlike EuO that requires an O$_2$ atmosphere, the growth of EuS avoids O$_2$ absorption on graphene and thus sidesteps the associated degradation effects. The EuS films have been proven insulating (see S1.3 in SI) and don't contribute to nonlocal signal in graphene/EuS heterostructure (see S2.5 in SI). Unless otherwise specified, all reported data were measured at 4.2 K using the AC lock-in method with an excitation current of 100 nA. The electrical and magneto-transport experiments were carried out in both the non-local and standard longitudinal/Hall configurations (Fig. 2a). In the non-local configuration, current $I$ is applied along leads 2 and 6 and voltage $V_{nl}$ is measured between leads 3 and 5. The nonlocal resistance is defined as $R_{nl} \equiv \frac{V_{nl}}{I}$. In the standard transport configuration, current $I$ is applied along 1 and 4, voltages $V_{xx}$ is measured between either 2 and 3 or 6 and 5, and longitudinal resistance is defined as $R_{xx} \equiv \frac{V_{xx}}{I}$. Under a perpendicular field $\mu_0 H$, the Hall voltages $V_{yx}$ is measured between either 6 and 2 or 5 and 3, and Hall resistance is defined as $R_{yx} \equiv \frac{V_{yx}}{I}$. We have measured > 20 control devices on > 10 different chips and ~ 15 surviving EuS/graphene devices on 5 different chips. (That fewer graphene/EuS chips are measured reflects the survival rate of graphene devices after EuS deposition - a result of electrostatic discharge either during deposition or in transit.) Compared to the control devices, the EuS-induced enhancement in R$_{nl}$ response ranges from unity to ~2 orders of magnitude. We estimate that the corresponding exchange field ranges from less than 1 Tesla to over 10 Tesla depending on the graphene and interface quality. The magnetization measurements were performed using a standard superconducting quantum interference device (SQUID) magnetometer.

**Figures and Captions:**

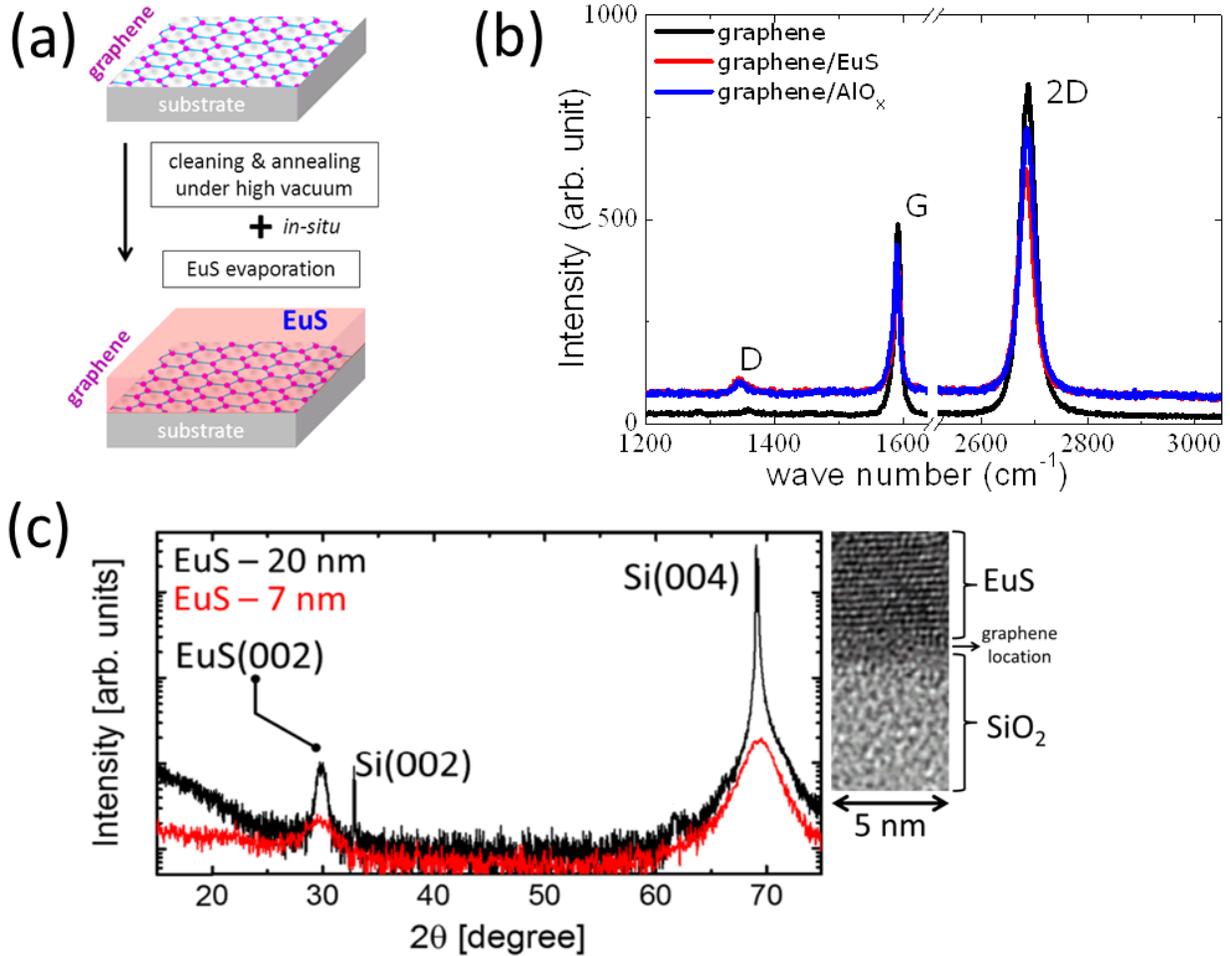

**Figure 1 (a)** Process for in-situ cleaning and deposition of graphene/EuS heterostructures under high vacuum (see Methods). **(b)** Comparison between the Raman spectroscopy data of pristine graphene (black line), graphene/EuS (red line) and graphene/AlO$_x$ (blue line). Both EuS and AlO$_x$ deposition preserve graphene's lattice structure. The characteristic graphene peaks *G* and *2D* with respect to the baseline show negligible changes. The defect *D* peak shows a slight increase in the spectral weight. **(c)** The wide angle X-ray diffraction (XRD) studies on our graphene/EuS heterostructure films with different EuS thickness (7 nm and 20 nm). Both EuS films show the same crystalline orientation with (002) diffraction peak, which confirms the high-quality, single-phase film growth.

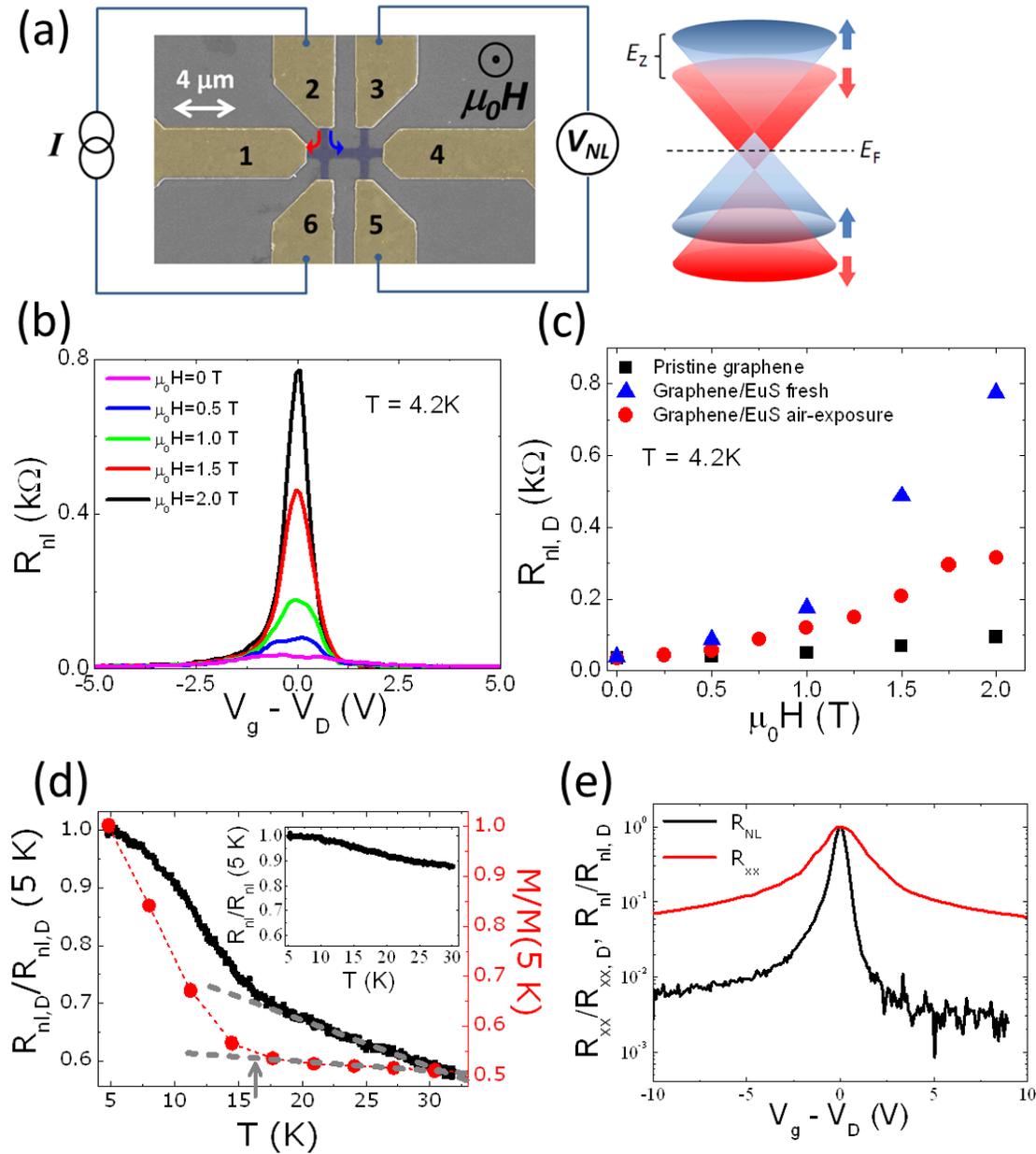

**Figure 2 (a) Left panel:** A false-colored device image taken by a scanning electron microscope (SEM). The central Hall bar region is graphene coated with EuS. The outer yellow regions are Ti/Pd/Au electrodes. Typical non-local measurements are carried out by applying current $I$ along leads 2 and 6 and measuring voltage $V_{nl}$ between leads 3 and 5. The nonlocal resistance is defined as $R_{nl} \equiv \frac{V_{nl}}{I}$. **Right panel:** Schematic drawing of the Zeeman splitting of the Dirac cone in graphene and the spin-up hole-like and spin-down electron-like carriers at the charge neutrality point. The applied field $\mu_0 H$ directs the oppositely spin-polarized charge carriers toward opposite directions along the Hall bar channel, generating a pure transverse spin current and namely the ZSHE. A nonlocal voltage drop is developed across the other pair of electrodes via the inverse effect. The flow of the spin-up (spin-down) current is shown by the blue (red) arrows in the SEM image. **(b)** Non-local resistance $R_{nl}$ as a function of gate voltage $V_g$ under different $\mu_0 H$ for a CVD-graphene/EuS device. At finite field, the large $R_{nl}$ peak near the Dirac point is the signature of ZSHE. At zero field, the $R_{nl}$ peak is very small, demonstrating negligible Ohmic contribution. **(c)** Comparison of $R_{nl,D}$ (the $R_{nl}$ value at the Dirac point $V_D$) vs. $\mu_0 H$ curves for the graphene device before (pristine) and after EuS deposition. Also plotted is the same graphene/EuS device after prolonged air exposure (~ 1 month), which shows that $R_{nl,D}$ strongly depends on the EuS quality. We note that the carrier mobility in graphene hardly varies after the air exposure (see S-1 in SI). **(d)** Comparison of $R_{nl,D}(T)$ (black line) and $M(T)$ (red circle) of the graphene/EuS heterostructure, confirming the magnetic origin of $R_{nl}$. Both data traces show an onset at ~ 16 K where EuS undergoes the ferromagnetic-paramagnetic transition. The $R_{nl,D}(T)$ dataset is taken at $\mu_0 H = 3.5$ T. The symbols represent actual data and the dashed lines are guide to the eyes. **Inset:** $R_{nl,D}(T)$ of graphene/AlO$_x$ at $\mu_0 H = 3.5$ T. The change in $R_{nl,D}$ is much weaker with no discernible onset feature. The gradual variation in $R_{nl,D}$ is consistent with the high-temperature background ($T > T_C$) of the graphene/EuS curve in the main panel. **(e)** Comparison of the normalized non-local resistance ($R_{nl}/R_{nl,D}$) and longitudinal resistance ($R_{xx}/R_{xx,D}$) at $\mu_0 H = 2.0$ T in graphene/EuS. $R_{nl,D}$ and $R_{xx,D}$ denote the resistance values at the Dirac point. The change in $R_{nl}/R_{nl,D}$ is one order of magnitude larger than $R_{xx}/R_{xx,D}$, thereby ruling out the Ohmic contribution as a primary source of the measured $R_{nl}$ under finite field.

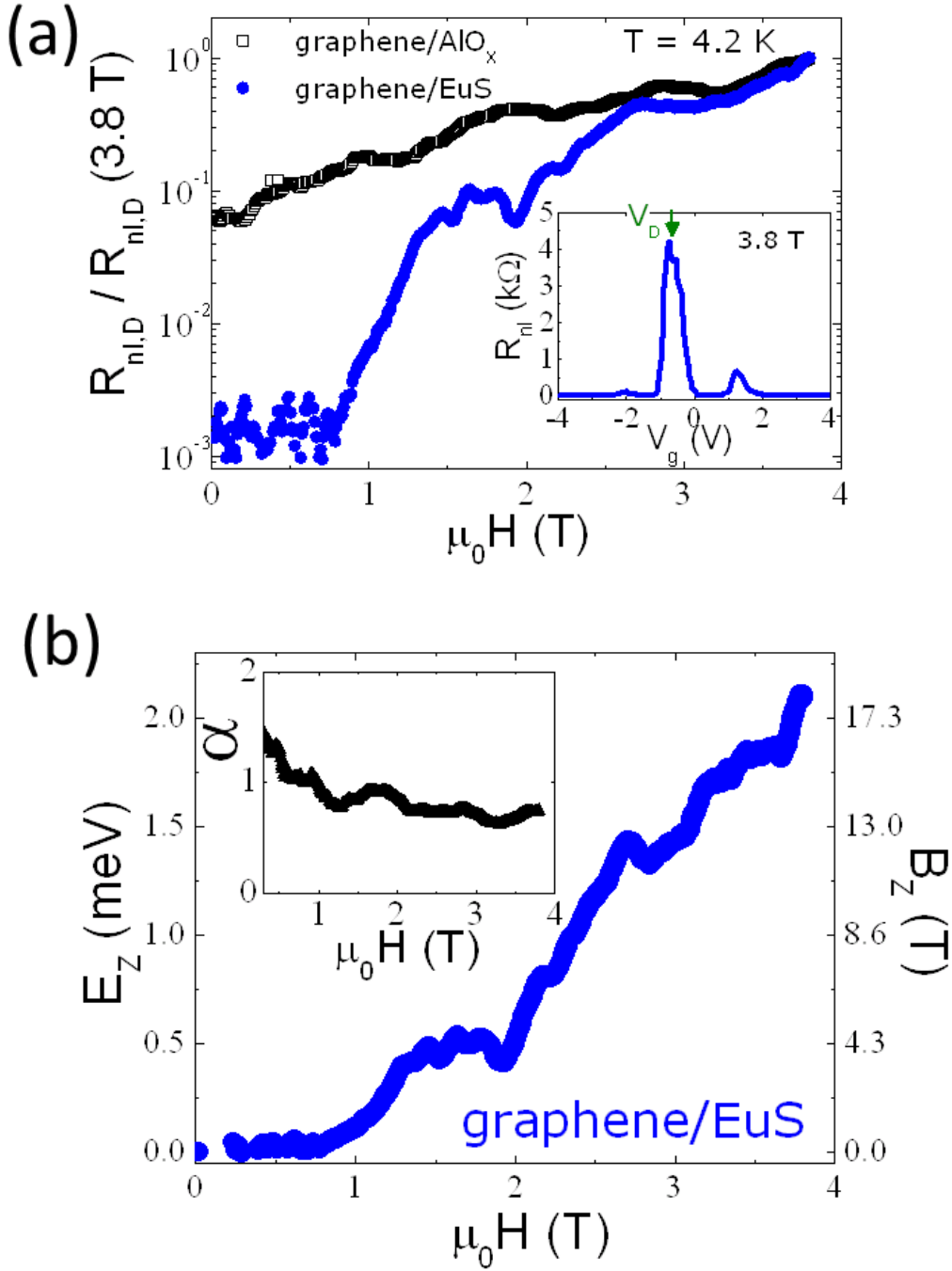

**Fig. 3 (a)** Field dependence of $R_{nl,D}$ in graphene/EuS vs. that in graphene/AlO$_x$, plotted in the form of normalized $R_{nl,D}/R_{nl,D}(3.8\ T)$. It shows orders of magnitude difference in the field-induced enhancement. The nonlocal resistance in graphene/AlO$_x$ (black square) increases gradually with the applied field. In contrast, the nonlocal resistance in graphene/EuS (blue circle) rises sharply at ~ 0.9 T. This onset behavior is commonly observed in graphene/EuS samples; however it is absent in pristine graphene or graphene/AlO$_x$ (e.g. Fig. 2c and Fig. 3a). We infer from the value of the estimated Zeeman splitting ($E_Z$) in Fig. 3b that the onset behavior develops when $E_Z$ is getting close to overcoming the thermal energy $k_B T$ (~ 0.36 meV at 4.2K), above which thermal smearing may no longer reduce the ZSHE. **Inset:** $R_{nl}$ vs. $V_g$ data of graphene/EuS at $\mu_0 H = 3.8\,T$, showing LL quantization and a pronounced EuS-induced ZSHE peak. **(b)** Quantitative estimation of the Zeeman splitting energy $E_Z$ in the presence of EuS. The curve represents the *lower-bound* estimate as elucidated in the main text. On top of the main curve, several secondary structures are seen. Since these features are history-dependent, irreversible upon thermal cycling and field cycling, we infer that these secondary features may be attributed to the multi-domain magnetization process of EuS, during which the spin current generation and transport may be modified by magnetic domains or domain walls. The right-hand-side axis shows the estimated magnetic exchange field exerted on graphene by EuS. **Inset:** Field dependence of the dimensionless coefficient $\alpha \equiv \sqrt{\frac{\beta(\mu_0 H)}{\beta(\mu_0 H = 1\,T)}}$ that captures the orbital-field effect in nonlocal resistance. $\alpha$ is extracted from the graphene/AlO$_x$ data in **(a)** where the interface exchange field is zero and the total Zeeman field equals the applied field.

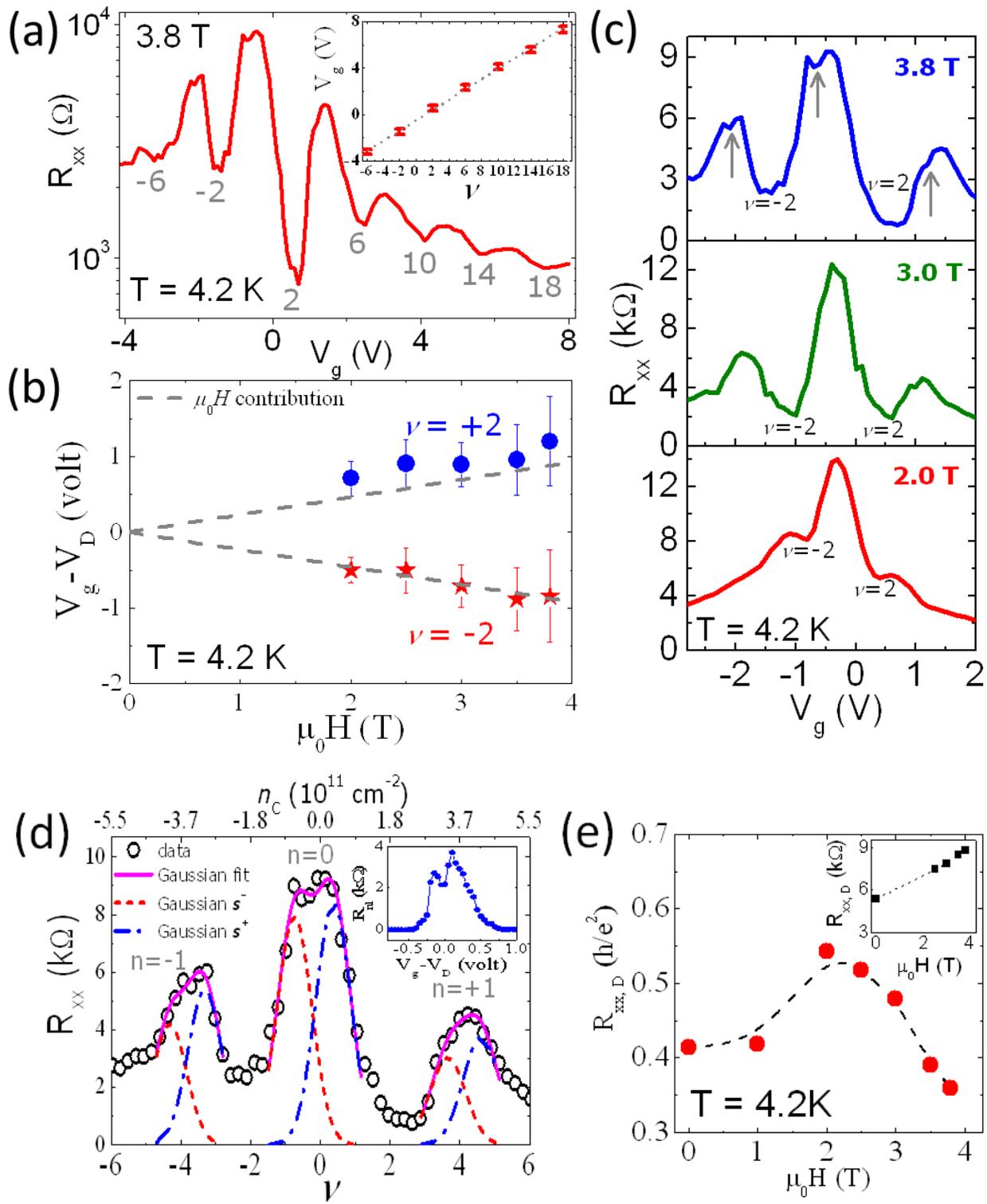

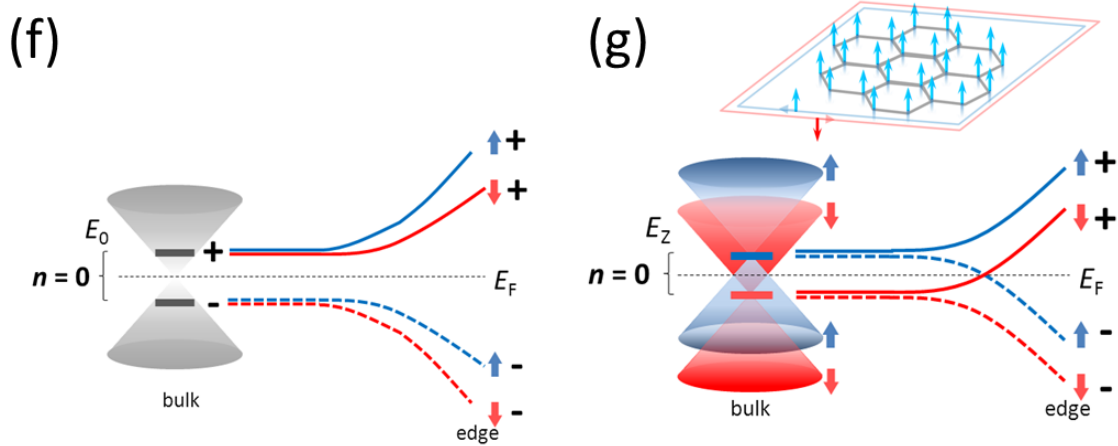

**Fig. 4 (a)** Magnetoresistance as a function of the gate voltage $V_g$ of the graphene/EuS device in Fig. 3, showing pronounced quantum oscillations at $\mu_0 H = 3.8$ T. The LL filling factors $\nu$ corresponding to each $R_{xx}$ minimum are labeled in the main panel. **Inset**: Gate voltage $V_g$ corresponding to each $\nu$. The dashed line is the linear fit to the data. From the intercept at $\nu = 0$, we determine the Dirac point $V_D$. **(b)** Field dependence of the gate voltage $(V_g - V_D)$ at $\nu = \pm 2$ extracted from the magnetoresistance curves. The dashed lines plot the calculated voltage values, taking into account only the applied field $\mu_0 H$ contribution. That the predicted $V_g$ matches well with the experimental data (symbols) within the error bar proves that the orbital effect induced by EuS is negligible. **(c)** $R_{xx}$ vs. $V_g$ data of graphene/EuS under different applied fields: 2.0 T, 3.0 T and 3.8 T. Quantum oscillations in $R_{xx}$ are observable down to $\mu_0 H = 2.0$ T. In contrast to the single peaks at 2 T, the main peak of $R_{xx}$ near the Dirac point develops double-peak features at 3.8 T, which suggests the splitting of the LL. Notably, the height of the $R_{xx}$ peak at the Dirac point decreases as $\mu_0 H$ increases. **(d)** Gaussian fit of the $R_{xx}$ vs. $V_g$ data at 3.8 T in (c) to quantify the LL splitting. The $x$-axis has been converted to $\nu$ according to the fitted values of $V_D$ and $\frac{dn}{dV_g}$ in (a). The splitting of the $n = \pm 1$ and $n = 0$ LLs is located at $\nu = \pm 4$, and $\nu = 0$ respectively, which indicates half filling of the $n = \pm 1$ and $n = 0$ LLs due to spin splitting. Simple Gaussians centered at each sub-peaks (sub-LLs): $R_{xx}^{peak} \propto \dfrac{1}{\gamma\sqrt{2\pi}} e^{-(V_g - V_0)^2 / 2\gamma^2}$ are used to identify the splitting. Here $\gamma$ describes the impurity broadening, and $V_0$ specifies the center of the split sub-peak. Each $n = 0$ and $\pm 1$ LLs is fitted by the superposition of two Gaussians with the same $\gamma$. The solid line shows the fitting curve with a Coefficient of Determination ~ 0.96, while the dotted and dash-dotted lines show the individual Gaussian components for each sub-peaks denoted as $s^+$ and $s^-$. **Inset**: The splitting of the n = 0 LL is also observed in the non-local resistance $R_{nl}$. (See Fig. S5-1(a) in SI.) **(e)** The resistance peak at the Dirac point $R_{xx,D}$ as a function of $\mu_0 H$. Symbols represent the actual data, and the dashed line is the guide to the eyes. $R_{xx,D}$ increases with $\mu_0 H$ at small field but decreases significantly above 2 T, in contrast to the diverging $R_{xx,D}$ in the valley-polarized spin-singlet $\nu = 0$ state shown in (f). Here we note that the overall peak height reduces, as can be seen in (c), and therefore the decrease of $R_{xx,D}$ does not come from the LL-splitting induced dip features. **Inset**: Monotonic increase of $R_{xx,D}$ vs. $\mu_0 H$ in the control sample graphene/AlO$_x$. **(f)** Schematic of the valley-polarized spin-singlet $\nu = 0$ state in which the bulk gap at the Dirac cone is dominated by the valley splitting $E_0$. The arrows indicate the spins. The "+" and "-" indicate different valleys. The resistance $R_{xx}$ diverges as the chemical potential is scanned across the charge neutrality point within the gap. **(g)** Schematic of the spin-polarized valley-singlet $\nu = 0$ state in which the bulk gap at the Dirac cone is dominated by the Zeeman splitting $E_Z$. The sub-LLs of the spin-up cone and spin-down cone crosses over in energy near the edge of the sample, leading to counter-propagating edge channels with opposite spins (see **Inset**) as in a quantum spin-Hall state. The presence of the edge channels agrees with the observation of decreasing $R_{xx,D}$ and $R_{xx,D}$ (3.8 T) < $R_{xx,D}$ (0 T) in (e). **Inset:** Schematic illustration of the vertically aligned spins in graphene by the EuS exchange field, leading to a ferromagnetic ground state.

**Acknowledgments**:

We thank J. Bucchignano, S. Dawes, B. Ek, J. Gonsalves, and Eileen Galligan at IBM for the technical support. P.W. and J.S.M. would like to acknowledge support from National Science Foundation Grant DMR-1207469, Office of Naval Research Grant N00014-13-1-0301, and the John Templeton Foundation Grant No. 39944. J.H. and S.L. would like to acknowledge support from Office of Naval Research Grant N00014-13-1-0662-01. D.H. would like to acknowledge support from the Nation Science Foundation Grant ECCS-1402378.

**Author Contribution**s:

C.T.C. conceived of the project. S.L., W.C. and J.H. fabricated the CVD graphene devices. P.W. and J.S.M. synthesized the graphene/EuS heterostructures. P.W., F.K., Y.Z. D.H., and J.S.M. characterized the EuS properties. F.L., S.L., L.P., W.C., D.C., and C.T.C. carried out the transport experiments. P.W. and C.T.C analyzed the data and drafted the manuscript. All authors commented on the manuscript.

# Supplementary Information:

## Strong Interfacial Exchange Field in 2D Material/Magnetic-Insulator Heterostructures: Graphene/EuS

**Authors:** Peng Wei[1,2,*], Sunwoo Lee[3,4], Florian Lemaitre[3,5,a], Lucas Pinel[3,5,b], Davide Cutaia[3,6,c], Wujoon Cha[7], Ferhat Katmis[1,2], Yu Zhu[3], Don Heiman[8], James Hone[7], Jagadeesh S. Moodera[1,2], Ching-Tzu Chen[3,†]

**Affiliations:**

1. Francis Bitter Magnet Laboratory, Massachusetts Institute of Technology, Cambridge, MA 02139, United States

2. Department of Physics, Massachusetts Institute of Technology, Cambridge, MA 02139, United States

3. IBM TJ Watson Research Center, Yorktown Heights, NY 10598, United States

4. Electrical Engineering Department, Columbia University, New York, NY 10027, United States

5. Institut polytechnique de Grenoble, F38031 Grenoble Cedex 1 - France

6. Politecnico di Torino, Turin 10129, Italy

7. Mechanical Engineering Department, Columbia University, New York, NY 10027, United States

8. Department of Physics, Northeastern University, Boston, MA 02115, United States

a. Present address: Department of Electrical Engineering, Eindhoven University of Technology, 5612 AZ, Eindhoven, The Netherlands

b. Present address: Department of Electronic & Electrical Engineering, The University of Sheffield, Mappin Street, Sheffield S1 3JD, United Kingdom

c. Present address: IBM Zurich Research Laboratory, Säumerstrasse 4, CH- 8803, Rüschlikon, Switzerland

* pwei@mit.edu

† cchen3@us.ibm.com

# S1. Material properties

## S1.1 Effect of EuS deposition on graphene mobility

We characterize the quality of the graphene channel before and after dielectric deposition (EuS or AlO$_x$) by the transconductance $g_m \equiv \frac{dI_{sd}}{dV_g}$ which is related to the field-effect mobility as: $\mu_{FE} = g_m l / V_{sd} w C_{ox}$, where $l$ ($w$) denotes the channel length (width), $V_{sd}$ denotes the source-drain voltage, and $C_{ox}$ denotes the back-gate capacitance.

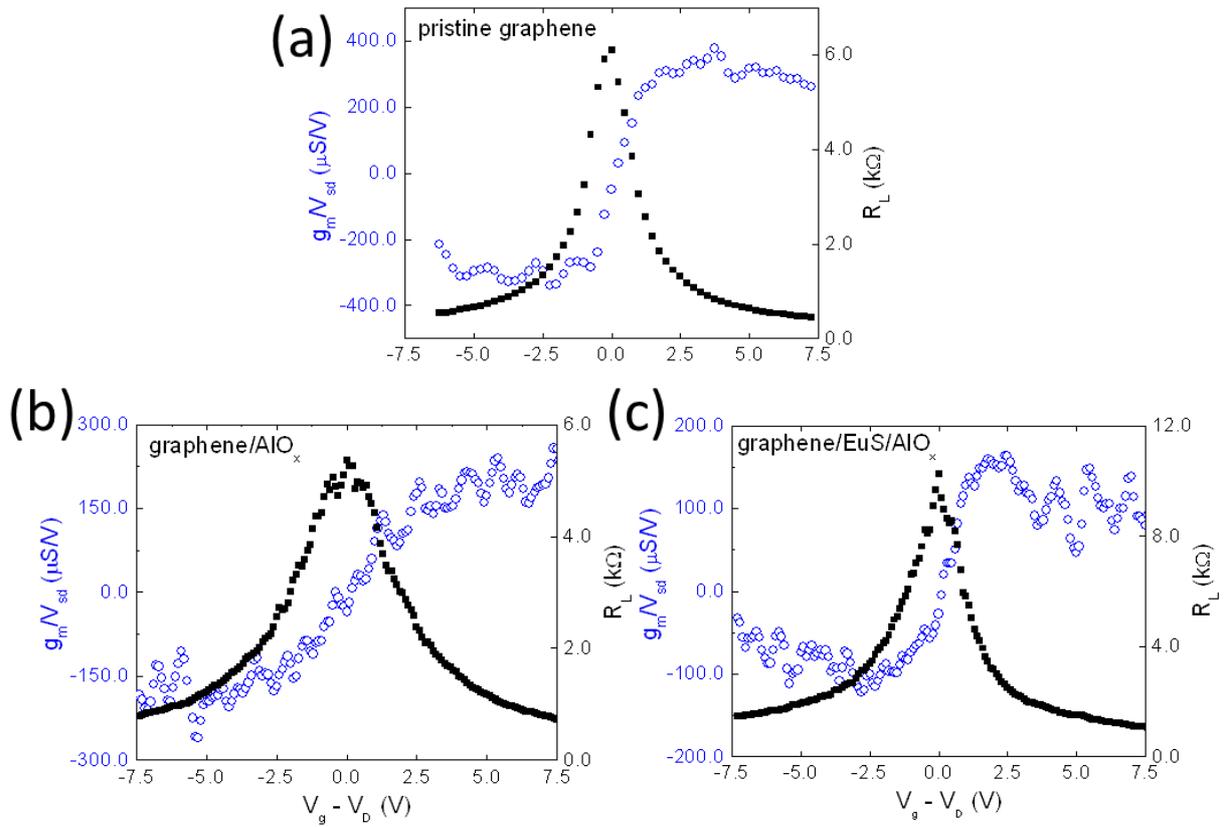

**Figure S1-1.** The $R_L$ vs. $V_g$ (black filled squares) and $g_m/V_{sd}$ vs. $V_g$ (blue hollow circle) data of **(a)** pristine CVD graphene; **(b)** CVD graphene/AlO$_x$; and **(c)** CVD graphene/EuS/AlO$_x$ device. The latter two devices have been studied in details in the main text. All three devices were fabricated from the same batch of CVD graphene.

Figure S1-1 shows the $R_L$ vs. $V_g$ (black filled square) and $g_m/V_{sd}$ vs. $V_g$ (blue hollow circle) data for a set of the CVD graphene devices in the pristine state (S1-1a) and with EuS/AlO$_x$ (S1-1c) or AlO$_x$ (S1-1b) overlayer. (The nonlocal resistance data of the

graphene/EuS/AlO$_x$ and graphene/AlO$_x$ devices are presented in the main text.) $R_L$ denotes the longitudinal resistance. These devices have been fabricated from the same batch of CVD graphene and patterned in the same process run. From the $g_m$ data, we see that the mobility of the graphene channel is generally reduced. Using a more accurate model that considers both the long- and short-range scattering, we fit the full-range gate voltage dependence of the conductivity[1] and deduce the following values of the mobility: ~ 16000 cm$^2$/Vs in pristine graphene, ~ 6000 cm$^2$/Vs in graphene/EuS and ~ 8000 cm$^2$/Vs in graphene/AlO$_x$. The decrease in mobility is consistent with the rise in the D-peak spectral weight in Raman spectra (see Fig. 1 in the main text).

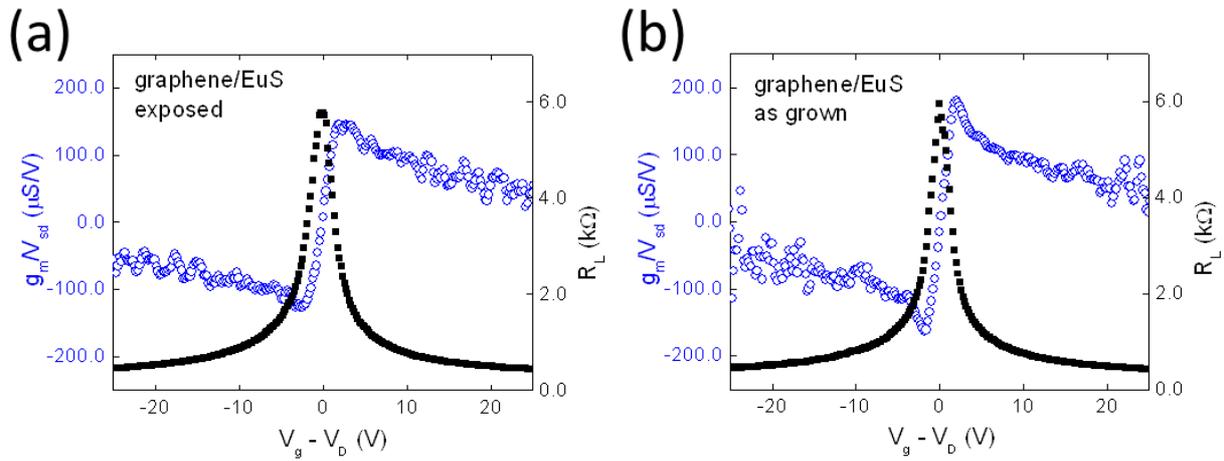

**Figure S1-2.** The $R_L$ vs. $V_g$ (black filled squares) and $g_m/V_D$ vs. $V_g$ (blue hollow circle) data of the CVD graphene/EuS/AlO$_x$ heterostructure device **(a)** before and **(b)** after prolonged air exposure, as discussed in Fig. 2d in the main text.

Figure S1-2 shows the mobility characterization data of the CVD-graphene/EuS device discussed in Fig. 2c in the main text, comparing the mobility of the device with freshly grown EuS vs. that after prolonged air exposure which partially oxidized EuS. The carrier mobility remains ~ 6000 cm$^2$/Vs despite the large reduction in nonlocal resistance, confirming that the source of nonlocal signal derives mainly from the magnetic exchange interaction between EuS and graphene.

We find that both the interface quality and the graphene quality play an important role in determining MEF. Without proper annealing, the residue on graphene suppresses the R$_{nl}$ enhancement. Besides, the higher the graphene quality before EuS deposition (manifested as higher mobility), the higher the probability is to observe significant $R_{nl}$ enhancement after EuS deposition.

## S1.2 Crystal structure and interface properties of graphene/EuS

Besides the XRD and TEM characterizations show in Fig. 1c of the main text, we further performed low-angle X-ray specular reflectivity (XRR) measurements to examine the graphene/EuS interface. XRR is a standard method that is sensitive to the interface roughness and long-range topological correlation.[2] From the period and decay of the oscillations in the reflectivity data (see Fig. S1-3), we confirm the thicknesses of EuS in graphene/EuS heterostructure. Furthermore, we deduce that the EuS-graphene interface to be atomically smooth (< 2Å, with less than 2% error in fitting calculation). Previous theoretical reports estimated an interatomic distance of 2.5 – 3 Å for generating a significant wave function overlap and exchange effect in graphene.[3,4] In order to accurately determine the interatomic distance at the interface, more advanced techniques, such as the synchrotron-based combined X-ray linear and circular dichroism, are required. However, our current characterizations have confirmed clean interface between graphene and EuS with smoothness within the range of exchange coupling.

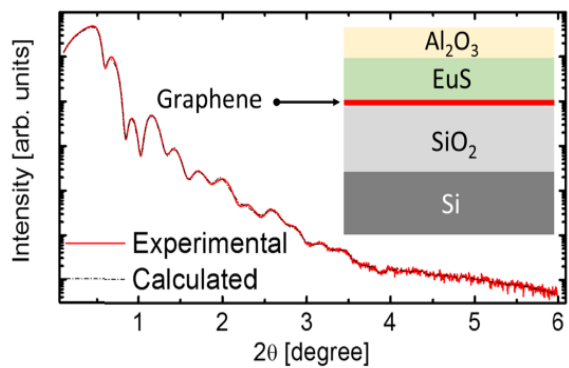

**Figure S1-3.** The standard low-angle X-ray specular reflectivity (XRR) studies for examining the interface roughness of graphene/EuS. The period and decay of the oscillations are related to the thickness of the thin film layers and their roughness.

## S1.3 The resistivity properties of EuS

To confirm the insulating nature of the EuS layer and its negligible contribution to the nonlocal resistance, we have conducted control experiments for the precise determination of the EuS resistivity. The EuS control samples have been grown under the same conditions as described in the manuscript. In contrast to other reports on conducting/semiconducting EuS samples with doping and impurities,[5,6] samples grown under this condition has been shown to be good insulators.[7] This type of insulating EuS thin films have been used as high quality tunnel barriers for a series of studies on the phenomena of spin-filter tunneling.[7]

In the control experiments, six EuS (7nm) Hall bar samples with an aspect ratio of l/w = 2 have been grown in high vacuum (low $10^{-8}$ torr) with in-situ evaporated metal contacts using shadow masks. All samples show a very large sheet resistance ~ $10^{11}$ Ω that corresponds to a resistivity ~ $10^4$ Ω·cm, which is much larger than the previously reported values in semiconducting EuS samples.[5] Compared to the sheet resistance of graphene (~ $10^4$ Ω at the Dirac point), the sheet resistance of EuS is ~7 orders of magnitude larger. Thus, for all practical purposes, EuS conduction is *negligible*. Furthermore, the resistance shows very weak temperature dependence (Fig. S1-4) – we don't see any anomaly in resistance near its Curie temperature.[5] Therefore, parallel conduction through EuS *cannot* explain the onset behavior in Fig. 2d of the manuscript.

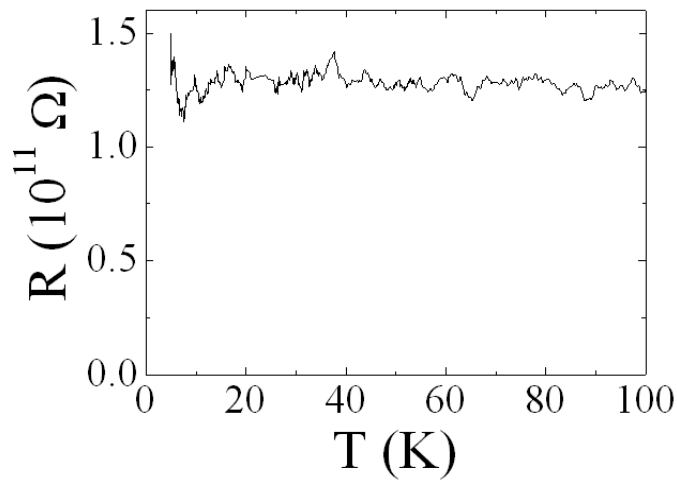

**Figure S1-4.** The temperature dependence of the resistance for a 7nm EuS thin film. No resistance anomaly is observed near its Curie temperature.[5]

## S2. Possible sources of $R_{nl}$

The non-local measurements in the configuration of $I$ (between leads 2-6) and $V_{nl}$ (between leads 3-5) may pick up erroneous voltage signals that are not due to ZSHE. In this section, several possible extrinsic sources that may contribute to finite $R_{nl}$ even without an external magnetic field are considered. However, they are found small in our experiment.

### S2.1 Ohmic contribution

The major erroneous voltage signal in our measured $R_{nl}$ is simply $R_{nl,\Omega}$ – the Ohmic contribution to the non-local resistance. $R_{nl,\Omega}$ is proportional to the resistivity $R_{xx}$ of the sample. For a Hall bar (Fig. 2a), the finite length/width ratio ($l/w$) of the channel will give rise to detectable voltage across leads 3-5 when a current is flowing across leads 2-6. In case of $l \geq w$, the Ohmic contribution is estimated as $R_{nl,\Omega} \sim R_{xx} \frac{w}{\pi l} \ln[\frac{\cosh(\frac{\pi l}{w})+1}{\cosh(\frac{\pi l}{w})-1}]$,[8] which leads to the relation $R_{nl,\Omega} \propto R_{xx}$. This Ohmic contribution exists even at $\mu_0 H = 0\,T$. Figure S2-1 compares $R_{nl}(\mu_0 H = 0\,T)$ with $R_{xx}(\mu_0 H = 0\,T)$ scaled by an factor $\sim \frac{1}{185}$. The excellent match between the two substantiates the Ohmic origin. Nevertheless under a finite field, say $\mu_0 H = 2.0\,T$, the peak value of $R_{xx} \leq 12\,k\Omega$ will only give $R_{nl,\Omega} \leq 65\,\Omega$, which is negligible compared to the total ZSHE peak signal ~ 800 $\Omega$. (See Fig. 2b and 2c in the main text.)

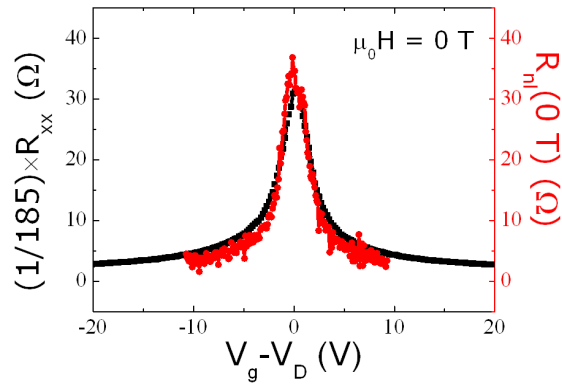

**Figure S2-1.** Comparison between the scaled resistance $R_{xx}$ and the non-local resistance $R_{nl}$ under zero applied field. They match very well with each other, which demonstrates that Ohmic contribution dominates the non-local signal at $\mu_0 H = 0\,T$.

## S2.2 Thermal contribution

In the high-field measurements that yielded the data shown in Fig. 3 and Fig. 4, the graphene heterostructure devices were immersed in liquid helium. Therefore, we don't expect to observe measurable thermal contribution to the nonlocal transport. This conjecture is corroborated by the experimental data as detailed below.

Joule heating and the Ettingshausen effects caused by the current *I* flowing along leads 2-6 may result in finite transverse voltage $V_{nl,H}$ across leads 3-5 under a perpendicular magnetic field.[9] This transverse voltage is a manifestation of the Nernst effect due to temperature gradient along the channel of the Hall bar. If *I* is an AC current with frequency *f*, as in the case of our experiment, Joule heating gives a transverse voltage that has 2*f* (i.e. *I*²) dependence $V_{nl}^{2f}$, while Ettingshausen effect gives a transverse voltage that has *f* dependence $V_{nl,E}^{f}$. For every sample, we carefully probe the 2*f* component of the non-local signal using lock-in technique. We find that $R_{nl}^{2f} < 5\ \Omega$ at 3.8 *T*, which is negligibly small compared to the $R_{nl}(3.8\ T)$ (see Fig. S2-2 below). Both Fig. S2-2a and b show that $R_{nl}^{2f}$ depends on the value of *I* (electrical current). *I* = 1.0 μA apparently results in larger $R_{nl}^{2f}$ compared to *I* = 0.1 μA. Furthermore, $R_{nl}^{2f}$ changes sign when flipping the current and voltage leads, which is due to the reversal of the heat flow along the channel. We note that the data presented in this paper are taken with *I* = 0.1 μA.

Ettingshausen effect causes a heat flow $\dot{Q}_E = S_{yx}TI$, where $S_{yx}$ is the Nernst signal, *T* is the temperature and *I* is the current along leads 2-6.[9] It thus builds up a temperature gradient along the channel: $\vec{\nabla}T \propto \dot{Q}_E = S_{yx}TI$, yielding a transverse Nernst voltage under a perpendicular field: $V_{nl,E}^{f} = S_{yx}\vec{\nabla}T \propto S_{yx}^2 TI$. Therefore, $R_{nl,E}^{f} \propto S_{yx}^2 T$ is the Ettingshausen contribution. Since $S_{yx}$ peaks at the Dirac point, $R_{nl,E}^{f}$ does, too. However, $R_{nl,E}^{f} \propto S_{yx}^2$ should also develops more beatings in quantum oscillations than $R_{xx}$ does in the quantum Hall regime. This is because, except for the *n* = 0 LL, $S_{yx}$ crosses zero and changes sign whenever the gate voltage $V_g$ is located in between adjacent LLs or in the middle of a Landau level (LL).[10-13] Our measured $R_{nl}^{f}$ clearly has only one peak at each *n* = ±1 LLs, similar to $R_{xx}$ (Fig. S2-3). The missing oscillations in the middle of the *n* = ±1 LLs contradicts the $R_{nl,E}^{f} \propto S_{yx}^2 T$ dependence. It therefore suggests that Ettingshausen contribution is negligible in our experiments.

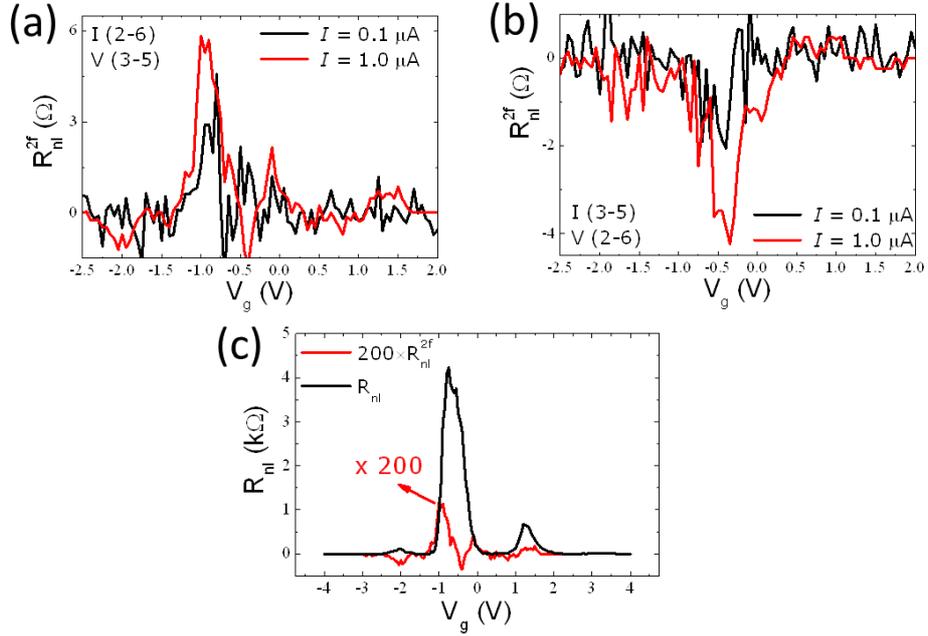

**Figure S2-2.** The $2f$ component of the non-local signal $R_{nl}^{2f}$ measured at $\mu_0 H = 3.8\ T$. It accounts for the Nernst voltages contributed by Joule heating. $R_{nl}^{2f}$ is as small as ~ 5 Ω. **(a)** and **(b)** demonstrate two measurement configurations with opposite heat flow directions for Joule heating, yielding opposite signs in the Nernst signal. **(c)** compares the $R_{nl}^{2f}$ signal (multiplied by 200) with the total signal of $R_{nl}$. $R_{nl}$ is orders of magnitudes larger, which confirms that Joule heating induced thermal contribution is negligible in our measurements.

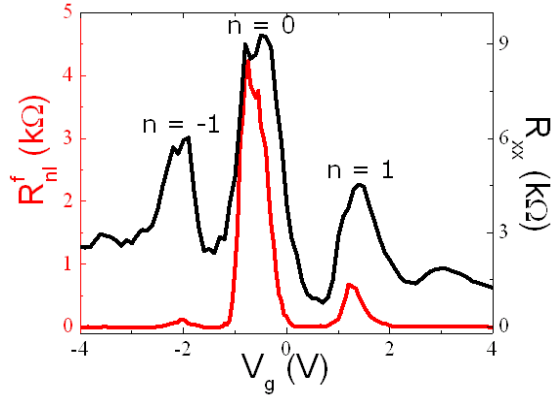

**Figure S2-3.** The comparison between the $1f$ component of the non-local signal $R_{nl}^{f}$ and the longitudinal resistance $R_{xx}$. No extra oscillations of $R_{nl}^{f}$ can be discerned at $n = \pm 1$ LLs, in contrary to the extra beatings in $S_{yx}^2$. This indicates that Ettingshausen effect related thermal contributions are negligible.

## S2.3 Spin-orbit effects

Two reports on CVD graphene samples and graphene in proximity to $WS_2$ have shown appreciable spin-orbit coupling and spin Hall effect, which are induced either by defects or by the proximity to materials with high spin-orbit coupling.[14,15] This results in an excess non-local resistance at *zero field* on top of the Ohmic signal.[15] In contrast, the *zero-field* nonlocal resistance of our CVD graphene/EuS devices scales very well with the longitudinal resistance (Fig. S2-1) and thus can be fully accounted for by the Ohmic contribution (see section S2.1), indicating that the proximity to EuS does *not* induce significant spin-orbit coupling.

The negligibly small spin-orbit effect is also confirmed by the observation of weak localization (Fig. S2-4) and the absence of anomalous Hall effect in the graphene/EuS heterostructures. In the small spin-orbit coupling limit, impurity scattering causes constructive interference between the time-reversal invariant scattering trajectories of the electrons; hence enhancing the transport resistivity at zero magnetic field, namely the weak localization behavior as shown in Fig. S2-4. In contrast, in the presence of substantial spin-orbit coupling, one would expect to see weak anti-localization instead.

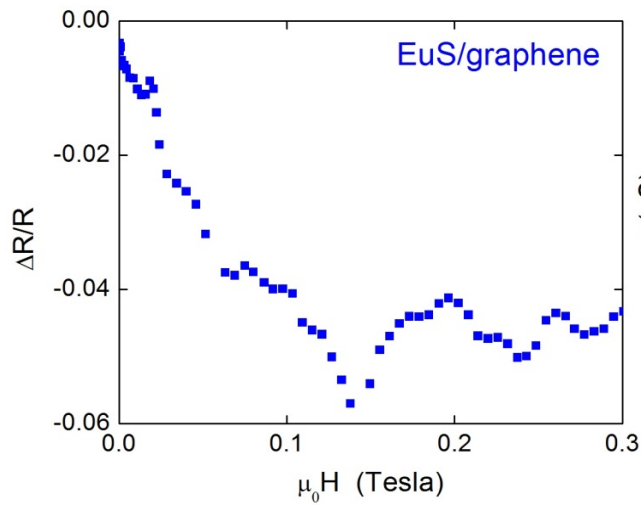

**Figure S2-4.** Low-field magnetotransport data of graphene/EuS. It shows the magnetoresistance change normalized to the zero field resistance. It demonstrates a sharp resistance peak at small field, which quickly drops down as the perpendicular magnetic field increases. The data is taken at 4.2 K.

## S2.4 EuS fringing field

The fringe field from the proximity to EuS, if sufficiently large, may generate an additional Lorentz force and contribute to the enhancement in $R_{nl}$. Furthermore, the stray field will cause measurable changes in the Landau level spacing. Comparing the quantum oscillation data of the magnetic graphene/EuS vs. that of the nonmagnetic graphene/ $AlO_x$ in Fig. S2-5, we see that the fringe field induced by EuS is negligible, since the charge carrier density per Landau level (the slope) is the same (within experimental errors) under the same applied field (3.8 T) in these two samples.

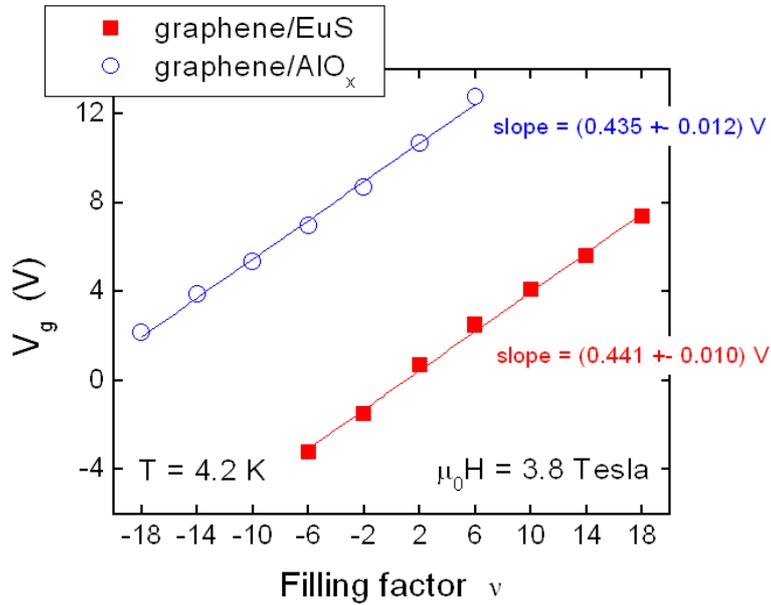

**Figure S2-5.** Back-gate voltage $V_g$ vs. filling factor $\nu$ for graphene/EuS (red filled square) and graphene/$AlO_x$ (blue hollow circle). The two samples were made from the same batch of CVD graphene, transferred onto substrates from the same Si wafer with an oxide thickness ~103 nm.

## S2.5 The non-local signal from EuS thin film

To verify that nonlocal transport through EuS is negligible, we have fabricated the EuS (7 nm) control devices with the exact same metal-contact configurations as in Fig. 2a of the manuscript. To mimic the experimental conditions, we apply a bias voltage (~ 1 meV) across one pair of the Hall-bar leads, measure the applied bias current I and the induced nonlocal voltage $V_{nl}$, and then derive the associated nonlocal resistance as $R_{nl} = V_{nl} / I$. At 4.2K, we measured the $R_{nl}$ of EuS at zero field and 2T field. The EuS $R_{nl}$ data show *no* field dependence and negligible gate voltage dependence (Fig. II-2), inconsistent with the data of graphene/EuS. The extremely large and negative $R_{nl}$ (>$10^{13}$ Ω) indicates that the voltage leads in the measurement setup are floating, similar to the behavior of an open-circuit device (without graphene and EuS) on the $SiO_x$/Si substrate. Thus, within the limit of our measurement electronics, EuS is as insulating as the back-gate dielectric $SiO_x$.

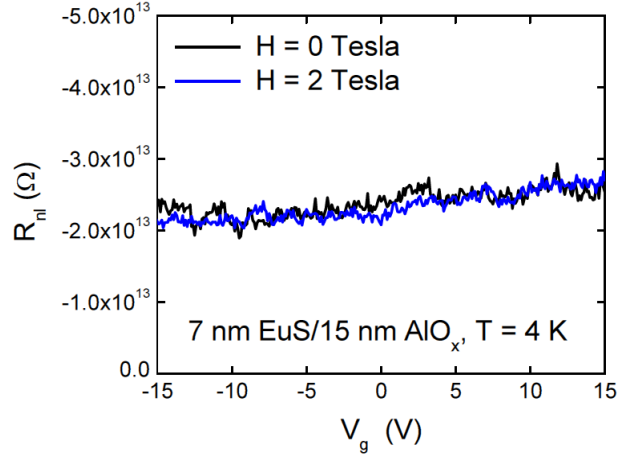

**Figure S2-6.** The nonlocal resistance of a 7nm EuS control devices with the same metal-contact configurations as in Fig. 2a of the manuscript. The data show *no* field dependence and negligible gate voltage dependence.

## S3. Approximated formula for ZSHE

It has been shown that spin Hall effect (SHE) can convert charge current into spin-mediated non-local charge transport over a scale comparable to the spin diffusion length.[8,16-18] Theoretically $R_{nl}$ is expressed as: $R_{nl} = \frac{V_{nl}}{I_{sd}} = \frac{1}{2}\theta_{SH}^2 \frac{w}{\sigma l_S} e^{-l/l_S}$, where the SHE coefficient or spin Hall angle is defined as $\theta_{SH} = \frac{\sigma_{SH}}{\sigma}$ with $\sigma_{SH}$ the spin-Hall conductivity and $\sigma = 1/\rho_{xx}$ the Ohmic conductivity.[16] The parameter $w$ ($l$) is the width (length) of the Hall bar channel, and $l_S$ is the spin diffusion length. In the case of ZSHE, spin Hall angle takes the form of $\theta_{SH} \propto \frac{\rho_{xy}^\uparrow}{\rho_{xx}^\uparrow} - \frac{\rho_{xy}^\downarrow}{\rho_{xx}^\downarrow} \approx E_Z \frac{\partial}{\partial \mu} \frac{\rho_{xy}}{\rho_{xx}}$, where $\mu$ is the chemical potential.[17] At the Dirac point, since $\frac{\partial \rho_{xx}}{\partial \mu} \approx 0$, the peak value of the non-local resistance is $R_{nl,D} \approx \frac{w}{2l_S} e^{-l/l_S} \frac{1}{\rho_{xx}} \left( E_Z \frac{\partial \rho_{xy}}{\partial \mu} \right)^2 \Big|_{\mu=\mu_D}$ with $E_Z$ the Zeeman splitting energy and $\mu_D$ the chemical potential at the Dirac point. This gives rise to Eq. 1 in the main text.

In Eq. 2 in the main text, we rewrite the peak value of the measured non-local resistance as $R_{nl,D} = R_0 + \beta(\mu_0 H) \cdot E_Z^2$ to include the extrinsic contribution $R_0$. The parameter $\beta$ thus takes the following form: $\beta = \frac{w}{2l_S} e^{-l/l_S} \frac{1}{\rho_{xx,D}} \left( \frac{\partial \rho_{xy}}{\partial \mu} \right)^2 \Big|_{\mu=\mu_D}$. In this equation, $\rho_{xx,D}$ and $\frac{\partial \rho_{xy}}{\partial \mu}\Big|_{\mu=\mu_D}$ depend on the magnetic field $B$ ($=\mu_0 H$), whereas $w$, $l$ and $l_S$ are field independent. The analytical formula of $\frac{\partial \rho_{xy}}{\partial n}\Big|_{n=0\,(\mu=\mu_D)}$ has a linear $B$ dependence given by Abanin et al. as: $\frac{\partial \rho_{xy}}{\partial n}\Big|_{n=0} = \frac{B}{4ecn_T^2}$ in the limit of $T \ll T_*$.[17] The parameters $n$ denotes the carrier density, $n_T = \frac{\pi}{12} \frac{k_B^2 T^2}{\hbar^2 v_F^2}$, $v_F$ is the Fermi velocity, and $T_* \sim 228$ K (the characteristic temperature above which the electron-hole drag dominates the electrical scattering – see the last paragraph in S3), well exceeding the measurement temperature $T = 4.2$ K. Hence in $\alpha = \sqrt{\frac{\beta(B)}{\beta(B_0)}}$ (where $B_0 = 1$ T as in the main text), the normalized value of $\frac{\partial \rho_{xy}}{\partial \mu} = \frac{\partial \rho_{xy}}{\partial n} \frac{\partial n}{\partial \mu}$ is sample *independent*.

The only sample-*dependent* term in $\alpha$ comes from $\rho_{xx,D}$, which is estimated as $\rho_{xx,D} = \frac{m_T}{2n_T e^2 \tau}(1+\tau^2 \varpi_c^2)$ in the limit of $T \ll T_*$. Here we take $m_T \approx 3.29 k_B T/v_F^2$ following Abanin et. al.,[17] $\tau$ is the scattering time, and $\varpi_{c,B} = \frac{eB}{m^*}$ is the cyclotron frequency. As a result, the sample-dependent term in $\alpha$ is $\sqrt{\frac{1+(\tau \varpi_{c,B_0})^2}{1+(\tau \varpi_{c,B})^2}}$.

The value of $\alpha$ does *not* vary much with the sample mobility. Recall that $\propto \sqrt{\frac{1+(\tau \varpi_{c,B_0})^2}{1+(\tau \varpi_{c,B})^2}} = \sqrt{\frac{1+(\mu_* B_0)^2}{1+(\mu_* B)^2}}$. The mobility $\mu_*$ of the samples is $\sim 6000$ cm$^2$/Vs in graphene/EuS and $\sim 8000$ cm$^2$/Vs for graphene/AlO$_x$ (see section S1.1). Even if there is a 25% difference between the mobility in graphene/AlO$_x$ and in graphene/EuS, the term $\sqrt{\frac{1+(\mu_* B_0)^2}{1+(\mu_* B)^2}}$ with $B_0 = 1$ T would only cause a $\sim 10\%$ correction to $\alpha$ at the highest field of the experiment (3.8 T).

In the aforementioned equations, the parameter $T_*$ determines the characteristic temperature above which the electron-hole drag dominates the electrical scattering. It is defined as $T_* = \gamma\sqrt{\hbar/\eta}$ where $\gamma = v_F\sqrt{e\hbar/\mu_*}$, and $\eta$ the electron-hole drag coefficient.[17] In the graphene/EuS samples, $\mu_* \sim 6000$ cm$^2$/Vs, and thus we estimate $\gamma \sim 346$ K by taking the standard value of $v_F \sim 9\times10^5$ m/s. We adopt $\eta \approx 2.3\hbar$ following Abanin et. al.,[17] since the dielectric constant of SiO$_x$ ($\varepsilon_0 = 3.9$) is similar to the value ($\varepsilon_0 \approx 4$) used in the reference. Therefore, $T_* \sim 228$ K, which is much higher than the measurement temperature 4.2 K.

## S4. Zeeman spin-Hall angle $\theta_{SH}$

We compute the interface exchange field enhanced effective spin Hall angle in graphene/EuS using the equations discussed above:

$$R_{nl,D} \approx R_0 + R_{ZSHE}, \text{ where } R_{ZSHE} = \frac{w}{2l_S}e^{-\frac{l}{l_S}}\frac{1}{\rho_{xx}}\left(E_Z\frac{\partial\rho_{xy}}{\partial\mu}\right)^2\bigg|_{\mu=\mu_D} \equiv \frac{1}{2}\theta_{SH}^2\frac{w}{\sigma l_S}e^{-l/l_S}. \quad (S1)$$

To estimate the spin diffusion length $l_S$, we compare the $R_{nl,D}$ values of two nearby devices with different channel length $l$ (2 μm vs. 3 μm) but the same channel width $w$ (1 μm). The ratio of $R_{ZSHE}$ is ~ (2.6 ± 1.1) at 2 T (see Fig. S4-1). The ratio of the sheet resistance is ~ 0.97, indicating good device homogeneity in electrical properties. Plugging these numbers in Eq. (S1), we get $l_S \approx (1.08 \pm 0.45)$ μm, and $\theta_{SH} \sim (0.51 \pm 0.10)$ at 2 T in pristine graphene.

After EuS deposition, $R_{nl,D}$ increases to (765 ± 25) Ω at 2 T and (3.91 ± 0.25) kΩ at 3.5 T. We may safely assume that $l_S \lesssim 1.08$ μm because the proximity to a ferromagnetic layer generally increases the spin-flip scattering rate in graphene. Since $R_{nl,D} \propto \theta_{SH}^2\frac{1}{l_S}e^{-\frac{l}{l_S}} \equiv \theta_{SH}^2 \cdot f(l_S)$ where $f(l_S)$ monotonically increases when $l_S < l$, we can derive the lower bound of $\theta_{SH}$ using the upper limit of $l_S \sim 1.08$ um. In the presence of the EuS interface exchange field, $\theta_{SH}$ has thus been enhanced to $\gtrsim 1.22$ at 2 Tesla and $\gtrsim 3.28$ at 3.5 Tesla.

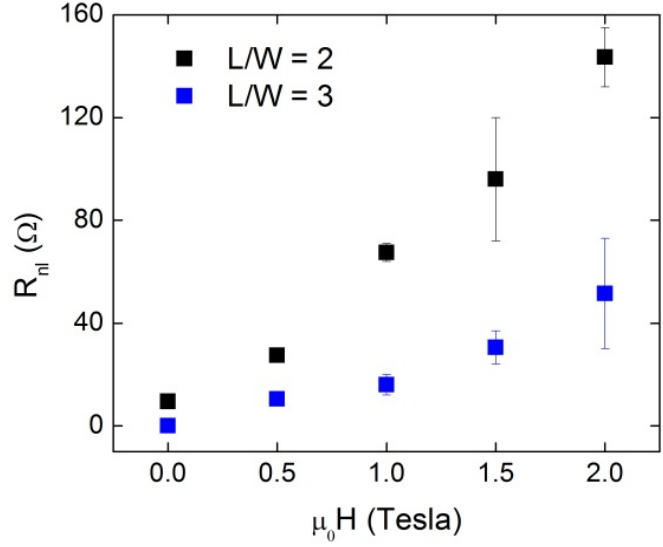

**Figure S4-1.** The comparison of the non-local resistance between two devices with different channel length/width ratio. The small $l/w$ enhances non-local resistance due to the term $e^{-\frac{l}{l_S}}$ in Eq. S1, which is used to estimate the spin diffusion length $l_S$.

## S5. Signatures of split LLs on other transport parameters

The splitting of the LLs can be also observed on other transport parameters, for example the Hall resistance $R_{xy}$, and non-local resistance $R_{nl}$. The $R_{nl}$ measurements are performed with finer sampling and ~10x longer averaging. We have also performed a comparison experiment with a 10x higher bias current (Fig. S5-1a bottom plot with $I = 1\mu A$) to suppress the measurement noises. The $R_{nl}$ signal shows a much smoother curve compared to the $I = 0.1\mu A$ data (Fig. S5-1a middle plot), but the splitting at the n = 0 LL is still clearly seen, further confirming its existence.

We further observe a plateau feature in $R_{xy}$ at the Dirac point that only exists at low temperature. Fig. S5-1b demonstrates two independent measurements of $R_{xy}$ vs. $V_g$ at $T = 4.7K$ and 16K under 3.5 T applied field. While both of them have quantized at ½ $e^2/h$, the plateau feature at the Dirac point is only observable at $T = 4.7K$. Fig. S5-1b insets show the derivative $dR_{xy}/dV_g$ for these two cases. While they have similar noise level, only the $T = 4.7K$ data shows the dip at the Dirac point, which corresponds to the development of the plateau in $R_{xy}$.

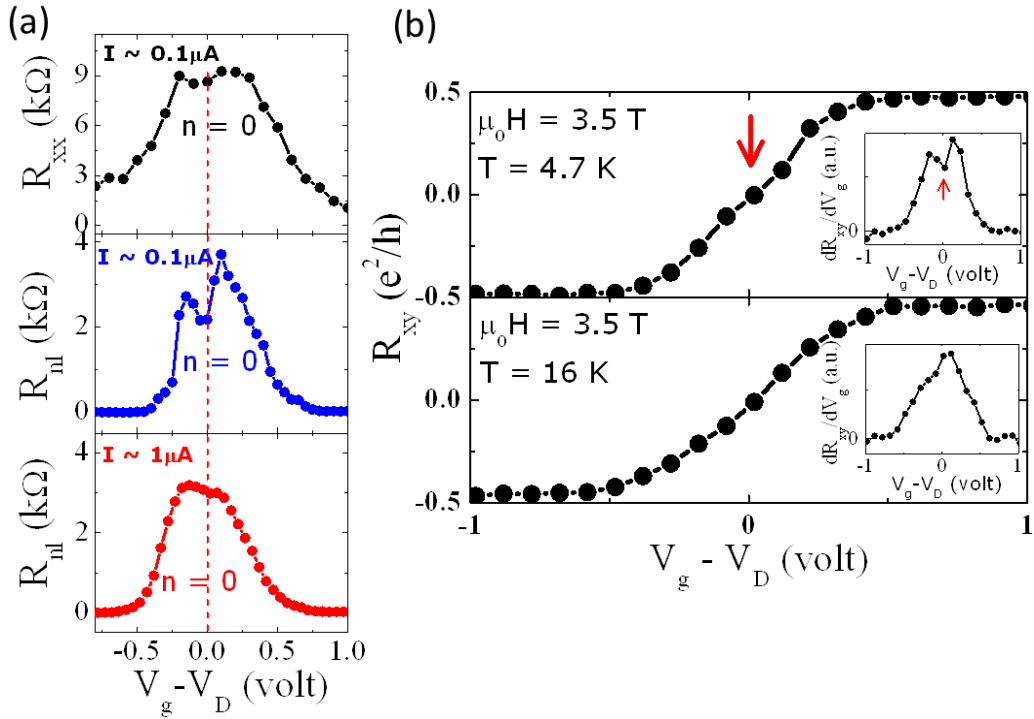

**Figure S5-1.** (a) The comparison of the splitting of the n = 0 LL for resistivity $R_{xx}$ (top), the nonlocal resistance $R_{nl}$ with $I \sim 0.1\mu A$ (middle), and $R_{nl}$ with $I \sim 1\mu A$ (bottom). All of them show the splitting feature. (b) The comparison of the $R_{xy}$ plateau at the Dirac point for $T = 4.7$ K and $T = 16$ K. The insets show the derivative of $R_{xy}$. The plateau feature only show at $T = 4.7$ K.